\documentstyle[preprint,aps,epsf]{revtex}  
\begin{document}
\draft



\title{Cross Sections for $\pi$- and $\rho$-induced Dissociation\\
 of $J/\psi$ and $\psi'$}

\author{Cheuk-Yin Wong$^1$, E. S. Swanson$^{2,3}$ and T. Barnes$^{1,4-6}$}

\address{$^1$Physics Division, Oak Ridge National Laboratory, Oak Ridge,
TN 37831 USA}

\address{$^2$Department of Physics and Astronomy,
University of
Pittsburgh, Pittsburgh, PA 15260 USA }

\address{$^3$Jefferson Lab, 12000 Jefferson Ave, Newport News, VA 23606 USA}

\address{$^4$Department of Physics, University of Tennessee, Knoxville, TN
37996 USA}

\address{$^5$Institut f\"ur Theoretische Kernphysik der Universit\"at
Bonn,  Bonn, D-53115, Germany} 

\address{$^6$Institut f\"ur Kernphysik, Forschungszentrum J\"ulich, 
J\"ulich, D-52425, Germany}

\maketitle

\begin{abstract}
{ We evaluate the cross sections for the dissociation of $J/\psi$ and
$\psi'$ by $\pi$ and $\rho$ at low collision energies, using the
quark-interchange model of Barnes and Swanson.  The dissociation cross
section for $J/\psi$ by $\pi$ is found to be relatively small with a
maximum of about 1 mb and a kinetic energy threshold of 0.65 GeV.  The
pion-induced $\psi'$ dissociation cross section is found to be much
larger, with a maximum of about 5 mb and a lower threshold.
Dissociation cross sections for $J/\psi$ and $\psi'$ by $\rho$ mesons
are also evaluated and are found to be large near threshold.  }
\end{abstract}

\newpage

The suggestion by Matsui and Satz \cite{Mat86} that $J/\psi$
production might be suppressed in a quark-gluon plasma has led to many
experimental and theoretical studies of $J/\psi$ production in
high-energy heavy-ion collisions.  The experimental observation by
NA50 \cite{Gon96,Rom98} of anomalous $J/\psi$ suppression in Pb+Pb
collisions in particular has been studied by many authors
\cite{Won96,Won98,Kha96,Bla96,Gav96,Cap96,Cas96,Sa99}.

The evolution of a $J/\psi$ or $\psi'$ produced in a heavy-ion
collision depends sensitively on charmonium dissociation cross
sections, which arise from processes such as low-energy inelastic
scattering of the $c\bar c$ state on $\pi$ or $\rho$ into open-charm
final states.  Small $J/\psi$ dissociation cross sections by $\pi$ or
$\rho$ may favor an interpretation of the Pb+Pb data in terms of the
production of a new phase of matter, possibly the quark-gluon plasma.
In contrast, a large $\rho$+$J/\psi$ dissociation cross section might
imply that $\rho$+$J/\psi$ inelastic scattering may be an important
part of the Pb+Pb anomaly because the density of $\rho$ mesons
increases approximately quadratically as the density of pions
increases.  In view of the importance of these dissociation cross
sections for the interpretation of heavy-ion collisions, they should
be evaluated and incorporated in Monte Carlo simulations before any
final conclusions are reached regarding the underlying physics.  

The dissociation of $J/\psi$ by light mesons has been considered
previously by several groups
\cite{Kha94,Kha96a,Mat98,Hag99,Mar95}. Unfortunately, the numerical
cross sections quoted in these references span a considerable range,
due largely to different assumptions regarding the dominant scattering
mechanism.

Kharzeev, Satz, and collaborators \cite{Kha94,Kha96a} used the parton
model and perturbative QCD ``short-distance'' approach of of Bhanot
and Peskin \cite{Pes79,Bha79}, and found remarkably small low-energy
cross sections for $J/\psi$ on light hadrons.  For example, their
$J/\psi$+$ N$ cross section at $\sqrt{s}=5$~GeV is only about $ 0.25\;
\mu b$ \cite{Kha94}.  A finite-mass correction increases this cross
section by about a factor of two \cite{Kha96a}.  However, in
high-energy heavy-ion reactions, the collisions between the produced
$\pi$ and $\rho$ with $J/\psi$ and $\psi'$ occur at low energies (of
the order of a few hundred MeV to about 1 GeV relative kinetic
energies). The applicability of the parton model and pQCD for
reactions at this low energy region is open to question.

Matinyan and M\"uller \cite{Mat98}, Haglin\cite{Hag99,Hag00}, and
Lin\cite{Lin99} recently reported results for these dissociation cross
sections in meson exchange models.  They use effective meson
Lagrangians and assume $t$-channel $D$ and $D^*$ meson exchange, which
leads to numerical results for $\pi$+$J/\psi$ and $\rho$+$J/\psi$
dissociation cross sections.  Matinyan and M\"uller found that these
cross sections are rather small; both are $\approx 0.2$-$0.3$~mb at
$\sqrt{s}=4$~GeV.  Including form factors (arbitrarily chosen to be
Gaussian with a width set to 1.5 GeV) would reduce the cross section
by an order of magnitude. Haglin obtained a very different result,
with much larger cross sections, by treating the $D^*$ and $\bar D^*$
mesons as non-Abelian gauge bosons in a minimally-coupled Yang-Mills
meson Lagrangian. Form factors were introduced in later calculations
and the mb-scale cross sections are sensitive to the choices of the
form factors \cite{Lin99,Hag00}.  Of course the use of a Yang-Mills
Lagrangian for charmed mesons has no {\it a priori} justification, so
the crucial initial assumption made in these references would require
independent confirmation.  In any case, the assumption of $t$-channel
exchange of a heavy meson such as a $D$ or $D^*$ between $\pi$ and
$J/\psi$ with point-like couplings is difficult to justify because the
range of these exchanges ($1/M \approx 0.1$ fm) is much smaller than
the physical sizes of the initial $\pi$ and $J/\psi$ mesons.

Charmonium dissociation processes can presumably be described in terms
of the fundamental quark and gluon interactions, but are of greatest
phenomenological interest at energy scales in the resonance region.
For this reason, we advocate the use of the known quark-gluon forces to
specify the underlying scattering amplitude, which must then be
convolved with explicit nonrelativistic quark model hadron
wavefunctions for the initial and final mesons.

Martins, Blaschke, and Quack \cite{Mar95} previously reported
dissociation cross section calculations using essentially the approach
we describe.  The short-distance interaction used by these authors in
particular is quite similar to the form we employ. For the confining
interaction, however, they used a simplified color-independent
Gaussian potential between quark-antiquark pairs only, rather than the
now well-established linear $\bbox {\lambda}(i) \cdot \bbox {
\lambda}(j)$ form.  They found a rather large $\pi$+$J/\psi$
dissociation cross section which reached a maximum of about 7 mb at
the kinetic energy in the center-of-mass system $E_{KE}$ of about 0.85
GeV.  Although our approach is very similar to that of Martins {\it et
al.}, our final numerical results differ significantly, due mainly to
the modeling of the confining interaction.

In this paper we use the approach discussed above to evaluate the
dissociation cross sections of $J/\psi$ by $\pi$ and $\rho$, and
compare our results to other theoretical cross sections reported in
the literature \cite{Kha94,Kha96a,Mat98,Hag99,Mar95}.  We also
calculate cross sections for the dissociation of $\psi'$ by $\pi$ and
$\rho$, which have not been evaluated elsewhere.

We employ the Barnes-Swanson quark-interchange model \cite{Bar92,ess}
to determine these dissociation amplitudes.  This approach uses the
nonrelativistic quark potential model and its interquark Hamiltonian
to describe hadron-hadron interactions and therefore implicitly
incorporates the successes of the quark model in describing the hadron
spectrum and many static properties of hadrons.  The model parameters
are fixed by fits to the meson spectrum, so there is little additional
freedom in determining scattering amplitudes and cross sections.  One
proceeds by calculating the scattering amplitude for a given process
at Born order in the interquark Hamiltonian. In the case of
meson-meson scattering, this scattering amplitude is given by the sum
of the four quark line diagrams shown in Fig. 1. These are evaluated as
overlap integrals of quark model wavefunctions, using the ``Feynman
rules" given in App. C of Ref.\cite{Bar92}.  This method has previously
been applied successfully to the closely related no-annihilation
scattering channels $I=2$ $\pi\pi$ \cite{Bar92}, $I=3/2$ $K\pi$ \cite
{Kpi}, $I=0,1$ S-wave $KN$ scattering \cite{KN} and the short-range
repulsive NN interaction \cite{NN}.

Following Ref.\cite{ess}, the interaction between each pair of
constituents $i$ and $j$ is taken to be
\begin{eqnarray}
\label{eq:Hij}
H_{ij}&=&{\bbox{\lambda}(i) \over 2}\cdot {\bbox{\lambda}(j) \over 2} \left \{
V_{\rm color Coulomb}(r_{ij})
+V_{\rm linear}(r_{ij})
+V_{\rm spin-spin}(r_{ij})
+V_{\rm con}
\right \}  \nonumber  \\
&=&{\bbox{\lambda}(i) \over 2}\cdot {\bbox{\lambda}(j) \over 2} \left \{
{\alpha_s \over r_{ij}} - {3\over 4} br_{ij}   
- {8 \pi \alpha_s \over
3 m_i m_j } \bbox{S}_i \cdot \bbox{S} _j \left ( {\sigma^3 \over
\pi^{3/2} } \right ) e^{-\sigma^2 r_{ij}^2}  
+V_{\rm con}
\right \}.
\end{eqnarray}
This Hamiltonian is derived in the Coulomb gauge, which is the most
convenient gauge for bound states and low-energy phenomena.  The model
parameter $\alpha_s$ is the strong coupling constant, $b$ is the
string tension, $m_i$ and $m_j$ are the interacting quark or antiquark
masses, and $\sigma$ is a range parameter in the Gaussian-smeared
spin-spin hyperfine interaction.  A constant shift $V_{con}$ is also
included in the interaction.  For antiquarks the generator
$\bbox{\lambda}/2$ is replaced by $-\bbox{\lambda}^{T}/2$.

\vskip 1cm

\vspace*{4.5cm}
\epsfxsize=300pt
\includegraphics{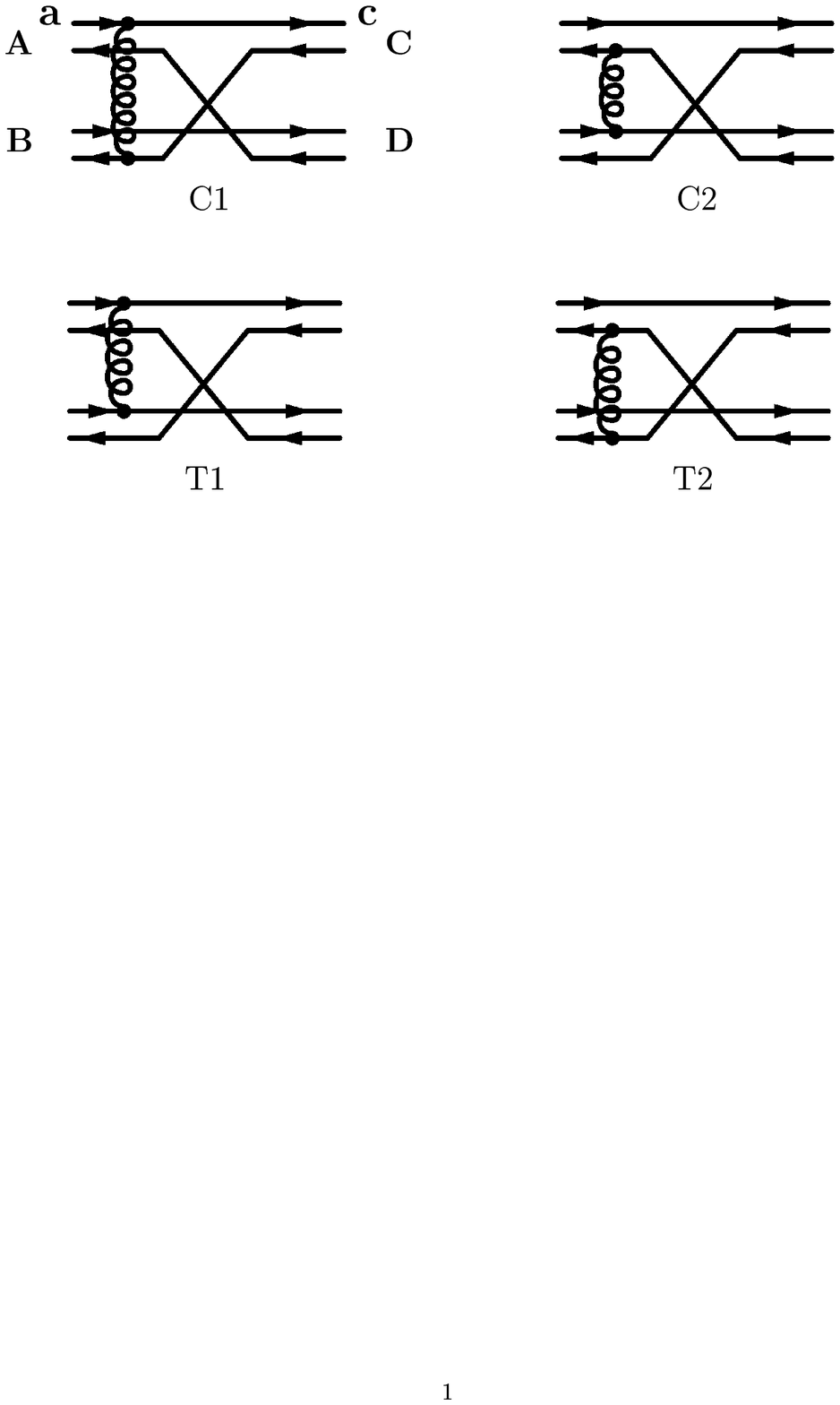}
\vspace*{0.4cm}\hspace*{4cm}
\begin{minipage}[t]{14cm}
\noindent {\bf Fig.\ 1}.  {Born-order quark line diagrams.\\ For
example, a specific channel is $A=J/\psi$, $B=\pi^+$,\\ $C=D^+$
and $D=\bar {D^*}$.}
\end{minipage}
\vskip 4truemm
\noindent 

\vskip 1cm

The model parameters we employed were $\alpha_s = 0.58$, $b = 0.18$
GeV$^2$, $\sigma = 0.897$ GeV, $m_u = m_d = 0.345$ GeV, $m_c = 1.931$
GeV and $V_{con} = -0.612$ GeV.  This set of parameters gives masses
within 0.08 GeV of experiment for the $\pi$, $\rho$, $D(1869)$,
$D^*(2010)$, $J/\psi$, and $\psi'$ mesons and also provides a very
good description of the $I=2$ S-wave $\pi\pi$ phase shift.  An
alternative set of parameters, found by fitting a large set of
experimental masses, is $\alpha_s = 0.594$, $b = 0.162$ GeV$^2$,
$\sigma = 0.897$ GeV, $m_u = m_d = 0.335$ GeV and $m_c = 1.6$~GeV.
This second set, with a flavor-dependent $V_{con}$, was used to test
the sensitivity of our results to parameter variations.

Before proceeding to our results, we note that the well-known
``post-prior ambiguity" arises in calculations of bound state
scattering amplitudes involving rearrangement collisions
\cite{Schiff}.  Since the Hamiltonian which describes the scattering
process $AB \rightarrow CD$ can be separated into free and interaction
parts in two ways, $H = H^{(0)}_A + H^{(0)}_B + V_{AB}$ or $H^{(0)}_C
+ H^{(0)}_D + V_{CD}$, there is an ambiguity in the choice of $V_{AB}$
or $V_{CD}$ as the interaction Hamiltonian.  The first version is
known as the ``prior" form and leads to the scattering diagrams of
Fig. 1, in which the interactions occur before quark interchange.  The
second choice is the ``post" form, which leads to diagrams in which
the interactions occur after quark interchange.  One may show that the
post and prior expressions for the scattering amplitude are equal,
provided that exact eigenfunctions of the free Hamiltonians are used
for the asymptotic states \cite{Schiff}.  (The relevance of this to
time reversal invariance is demonstrated numerically in
Ref.\cite{ess}.)  In our calculations we employ numerically determined
Hamiltonian eigenfunctions for each of the external meson states
considered; in the nonrelativistic case this would suffice to
eliminate the post-prior discrepancy.  In the processes considered
here we have used relativistic kinematics and phase space, but use
Galilean boosts for the states, as appropriate for a nonrelativistic
quark model calculation.  In consequence we find that the post and
prior scattering amplitudes differ slightly.  (We note in passing that
one could carry out a relativised version of this calculation,
although the full relativistic boosts would induce small additional
effects due to Wigner rotations and creation of quarks and gluons.)
In this paper we use the mean of the post and prior results as our
theoretical cross section, and the estimated errors due to the
post-prior discrepancy and parameter variations are indicated by bands
in the figures.

The cross sections we obtain for the dissociation of $J/\psi$ and
$\psi'$ by $\pi$ are shown in Fig.~2 as a function of the kinetic
energy in the center of mass system, $E_{KE} = \sqrt{s}-M_A-M_B$,
where $M_A$ and $M_B$ are the rest masses of the colliding particles
in the initial channel.  The lowest-lying allowed final states are
$\bar D^* D$, ${\bar D}D^*$, and $\bar D^*D^*$, and the total
dissociation cross section is taken to be the sum of these three
channel cross sections; this is shown as a solid line in the figure.
( The reactions $\pi$+$J/\psi \to \bar D D$ and $\pi$+$\psi' \to \bar
D D$ are $\Delta S \neq 0$ transitions allowed in QCD but have
zero transition matrix elements in our Hamiltonian (1). These
transition amplitudes would be nonzero for example if we included
spin-orbit terms in (1).  The relatively weak process $\pi$+$J/\psi
\to \bar D D$ has been considered in a Dyson-Schwinger formalism by
Blaschke $et~al.$ \cite{Bla00}, who find a maximum cross section of
about 0.1~mb near threshold.  Note that the S-wave to S-wave transition
is absolutely forbidden, so although $\pi$+$\psi' \to \bar D D$ is
actually exothermic, it does not lead to a divergent cross section at
threshold.)

The $\pi$+$J/\psi$ dissociation process is endothermic and requires an
initial kinetic energy of 0.65~GeV. The cross section shows a rapid
rise above threshold (as expected for an S-wave process) and has a
broad maximum of about 1 mb not far above threshold (Fig. 2$a$). This
is somewhat smaller than the $\approx 7$~mb estimated by Martins
$et~al.$, which we discuss below.
 
\vspace*{-1.0cm}
\epsfxsize=300pt
\includegraphics{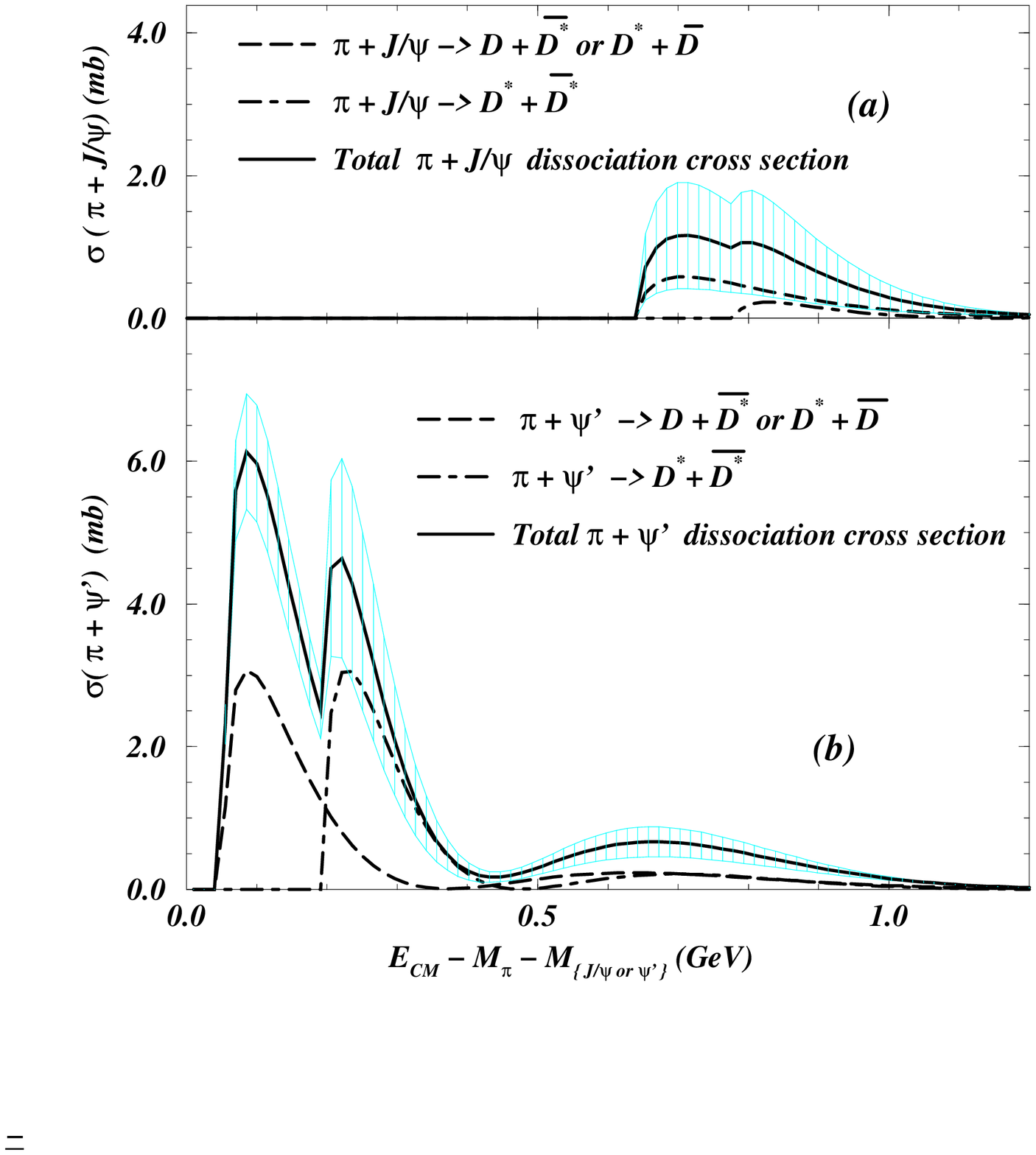}
\vspace*{9.8cm}\hspace*{1.5cm}
\begin{minipage}[t]{14cm}
\noindent {Fig. 2.  Cross sections for pion-induced dissociation of
$J/\psi$ (Fig. 2$a$) and $\psi'$ (Fig. 2$b$).  The solid curves give
the total dissociation cross section.  Estimated systematic errors due
to parameter uncertainties and the post-prior discrepancy are shown as
bands.}
\end{minipage}
\vspace*{0.5cm}

The cross section for dissociation of the $\psi'$ by $\pi$ is rather
larger in part because this reaction is only weakly endothermic; the
initial $\pi$+$\psi'$ kinetic energy in $\pi$+$\psi'\to\ \bar D D^*$
and $D \bar D^*\ $ is only about 0.05~GeV at threshold.  The total
cross section reaches a maximum of about 6.2(0.8)~mb at the kinetic
energy of about 0.1~GeV and has a secondary maximum of 4.6(1.8)~mb at
the kinetic energy of about $0.22$~GeV due to the opening of the
$D^*\bar D^*$ channel.  Notice that the ratio of the peak values of
the $\pi$+$\psi'$ and $\pi$+$J/\psi$ cross sections is roughly 6;
this should be contrasted with the prediction of $\sim 5000$ given in
Ref. [17]. The minimum in the cross section near the kinetic energy of
0.4 GeV is due to the complete destructive interference between
transfer (T1 and T2) and capture (C1 and C2) diagrams.

We next calculate the $\rho$+$J/\psi$ and $\rho$+$\psi'$ dissociation
cross sections.  The allowed low-lying final states are $D \bar D$, $D
\bar D^*$ and $D^* \bar D$ ($S_{tot} = 0,1$), and $D^* \bar D^*$
($S_{tot} = 0,1,2$). These cross sections are shown in Fig. 3. Since
the reaction $\rho+J/\psi \rightarrow D \bar D$ is exothermic, this
cross section diverges as $1/|\vec v_{\rho J/\psi }|$ near threshold.
For other channels the thresholds occur at higher energies, so those
subprocesses are endothermic.  The total dissociation cross section is
shown as a solid line in Fig. 3.  It is numerically about 11(3) mb at
a kinetic energy of 0.1 GeV, decreasing to 6(2) mb at a kinetic energy
of 0.2 GeV.

\vspace*{-1.0cm}
\epsfxsize=300pt
\includegraphics{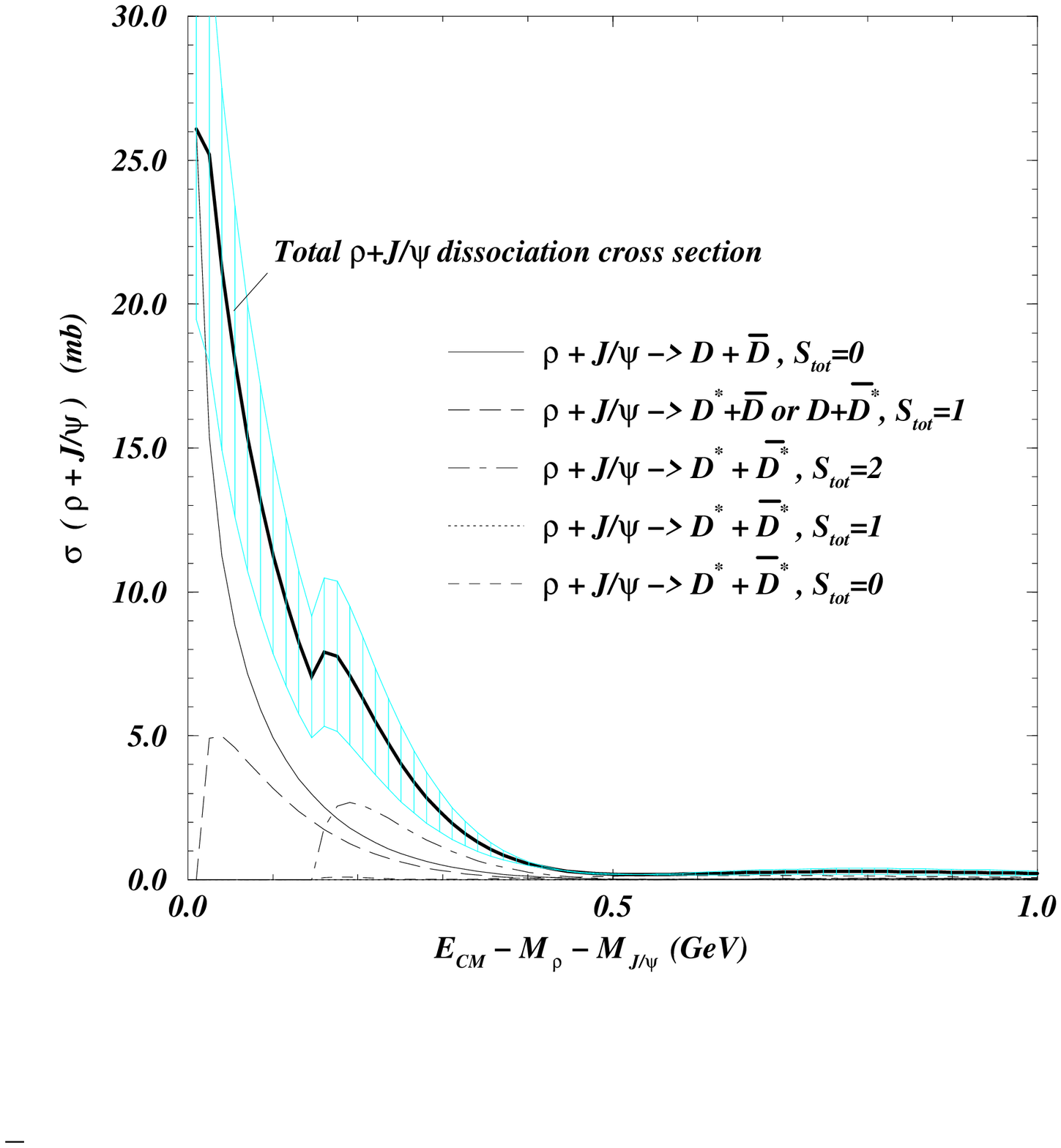}
\vspace*{9.6cm}\hspace*{1.5cm}
\begin{minipage}[t]{14cm}
\noindent {\bf Fig.\ 3}.  {
Total and individual channel cross sections 
$J/\psi$ dissociation by $\rho$.
}
\end{minipage}
\vskip 4truemm
\noindent 
\vspace*{-1.2cm}

In the case of $\rho$+$\psi'$ dissociation, all the channels we consider
are exothermic, so the low-energy divergence is quite pronounced.  Our
numerical results for these cross sections are shown in Fig. 4.  The
total cross section decreases from 15(2) mb to 6(2) mb as the kinetic
energy increases from 0.1 to 0.2 GeV.

\vspace*{-0.9cm}
\epsfxsize=300pt
\includegraphics{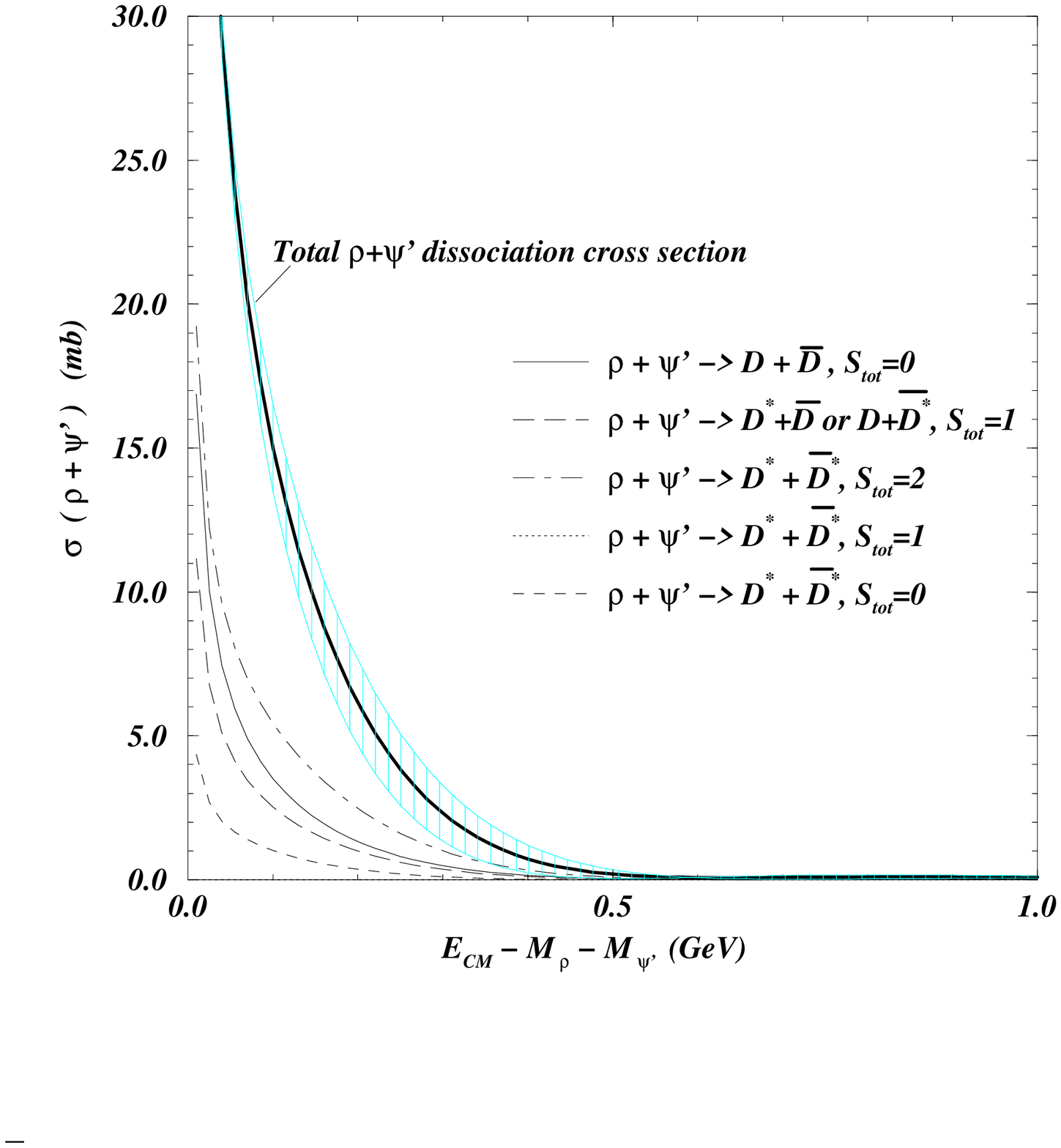}
\vspace*{9.6cm}\hspace*{1.5cm}
\begin{minipage}[t]{14cm}
\noindent {\bf Fig. 4}.  {
Total and individual channel cross sections 
for $\psi'$ dissociation by $\rho$.
}
\end{minipage}
\vskip 4truemm
\noindent 
\vspace*{-1.2cm}

We previously noted that our $\pi$+$J/\psi$ cross section is
considerably smaller than the estimate of Ref.\cite{Mar95}, although
we use a similar approach.  There are several differences between the
two approaches which lead to this discrepancy.  Martins {\it et al.}
assumed that the confining interaction is an attractive Gaussian
potential which acts only between quark-antiquark pairs.  The neglect
of the quark-quark and antiquark-antiquark confining interaction
amounts to discarding the transfer diagrams (T1 and T2) for the
confining potential. Since we find that the transfer and capture
diagram confinement contributions are similar in magnitude but
opposite in sign (due to color factors), Martins {\it et al.}  did not
include an important destructive interference.  Their use of a
Gaussian rather than a linear potential will obviously lead to
quantitatively different cross sections.  Furthermore, the cross
section values are quite sensitive to the parameters used when T1 and
T2 are not included.  All these factors contribute to the differences
between our results and those of Martin $et~al.$ for $\pi$+$J/\psi$
collisions.

The destructive interference between transfer and capture diagrams
with spin-independent forces (color Coulomb and confinement) has been
noted previously (see, for example, Refs.\cite{Bar92,ess} and
references cited in \cite{NN}).  This interference explains the
well-known spin-spin hyperfine dominance in light hadron scattering in
channels such as $I=2$ $\pi\pi$ and $NN$. In the presence of heavy
quarks, however, most hyperfine interaction diagrams are suppressed by
the large charm quark mass; this is the reason we included the color
Coulomb and confining interactions in the present analysis.  It is
interesting in retrospect to examine the different channels and
determine which of the various interactions dominates the amplitude.
We find that the hyperfine interaction still dominates the
$\pi$+$J/\psi$ dissociation amplitude, whereas the linear confining
interaction dominates $\pi$+$\psi'$, $\rho$+$\psi$, and (for the
kinetic energy less than 0.3 GeV) $\rho$+$\psi'$.  Above $0.3$ GeV,
the color Coulomb interaction dominates $\rho$+$\psi'$ scattering.
(We caution the reader that this decomposition depends on the choice
of the post or prior form for the T-matrix; the results we quote are
for the prior form, involving the diagrams of Fig. 1.)

There is no direct experimental measurement of these cross sections to
which we can compare our results.  We found a small $\pi$+$J/\psi$
cross section which starts at a high threshold and a large
$\pi$+$\psi'$ cross section which starts at a low threshold.  If these
cross sections are folded with a distribution of pions with an average
kinetic energy of about 200 MeV, we would obtain $\sigma_{\rm
effective} (\pi+J/\psi) << \sigma_{\rm effective}(\pi+\psi')$, which
is consistent with earlier observation in a model of $J/\psi$ and
$\psi'$ suppression in O+A, and S+U collisions \cite{Won96}.
Hopefully, future Monte Carlo simulations of the dynamics of
charmonium in heavy-ion collisions will lead to a more direct
comparison.  The large $\rho$+$J/\psi$ and $\rho$+$\psi'$ cross
sections we have found imply that both the $J/\psi$ and $\psi'$ will
be quickly dissociated if there is a significant $\rho$ meson
population in the medium.  Since our results on the divergence of
$\rho$+$J/\psi$ and $\rho$+$\psi'$ dissociation cross sections at
threshold follow directly from simple kinematics, these results must
be qualitatively correct. The normalization of these cross sections,
however, required detailed calculation and should also be compared to
experiment if possible.  Since dissociations of $J/\psi$ by $\pi$ and
$\rho$ populate different states (for example $\pi$+$J/\psi$ doe not
lead to $D \bar D$ but $\rho$+$J/\psi$ does), it may be possible to
separate these processes and their associated cross sections by
studying the relative production of $D\bar D$, $D^*\bar D$+$h.c.$ and
$D^*\bar D^*$ if the expected open charm background can be subtracted.

In the future it may be useful to carry out detailed simulations of
$J/\psi$ absorption in heavy-ion collisions using the cross sections
obtained here to test the accuracy of our results. If our cross
sections do prove to be reasonably accurate, it will clearly be useful
to incorporate them in simulations of $J/\psi$ suppression in Pb+Pb
collisions and in other processes that use charmonium as a signature
of the quark-gluon plasma in order to subtract the effects of $J/\psi$
suppression due to its interaction with hadron matter.

\vspace*{-0.5cm}
\section*{Acknowledgments}
\vspace {-0.5cm}

This research was supported by the Division of Nuclear Physics, DOE,
under Contract No. DE-AC05-96OR21400 managed by Lockheed Martin Energy
Research Corp. ES acknowledges support from the DOE under grant
DE-FG02-96ER40944 and DOE contract DE-AC05-84ER40150 under which the
Southeastern Universities Research Association operates the Thomas
Jefferson National Accelerator Facility. TB acknowledges additional
support from the Deutsche Forschungsgemeinschaft DFG under contract Bo
56/153-1.  The authors would also like to thank D. B. Blaschke,
C. M. Ko, G. R\"opke and S. Sorensen for useful discussions.

\end{document}